\documentclass[prb, twocolumn, aps,amsmath,amssymb,amsfonts, superscriptaddress]{revtex4}
\usepackage[german,american,english]{babel}
\usepackage{graphicx}
\usepackage{graphics}
\usepackage{dcolumn}
\usepackage{multirow}
\usepackage{bm}
\usepackage{latexsym}
\usepackage{amssymb}
\usepackage{amsmath}
\usepackage{amsfonts}
\usepackage{layout}
\usepackage{verbatim}
\usepackage{epsfig}
\usepackage{graphicx}
\usepackage{amsbsy}
\usepackage{lipsum}

\newcommand{\bea}{\begin{eqnarray*}}
	\newcommand{\eea}{\end{eqnarray*}}
\newcommand{\bne}{\begin{equation*}}
\newcommand{\ede}{\end{equation*}}

\newcommand{\bnen}{\begin{equation}}
\newcommand{\eden}{\end{equation}}
\newcommand{\bean}{\begin{eqnarray}}
\newcommand{\eean}{\end{eqnarray}}
\newcommand{\bsen}{\begin{subequations}}
	\newcommand{\esen}{\end{subequations}}

\newcommand{\bna}{\begin{array}}
	\newcommand{\eda}{\end{array}}
\newcommand{\bnm}{\begin{enumerate}}
	\newcommand{\edm}{\end{enumerate}}

\newcommand {\bkt} [1] {\langle #1 \rangle}

\newcommand {\pd} [2] {\frac{\partial #1}{\partial #2}}

\usepackage[utf8]{inputenc}

\begin{document}

\title{Semiclassical equations of motion for disordered conductors: extrinsic interband velocity, corrected collision integral and spin-orbit torques}
\author{Rhonald Burgos Atencia}
\affiliation{
Facultad de Ingenier\'ias, Departamento de Ciencias B\'asicas,
Universidad del Sin\'u, Cra.1{\rm w} No.38-153, 4536534,
Monter\'ia, C\'ordoba 230002, Colombia}
\author{Qian Niu}
\affiliation{Department of Physics, The University of Texas at Austin, Austin TX 78712}
\author{Dimitrie Culcer}
\affiliation{School of Physics, The University of New South Wales, Sydney 2052, Australia}
\begin{abstract}
The semiclassical equations of motion are widely used to describe carrier transport in conducting materials. Nevertheless, the substantial challenge of incorporating disorder systematically into the semiclassical model persists, leading to quantitative inaccuracies and occasionally erroneous predictions for the expectation values of physical observables. To address this issue, in the present work we provide a general prescription for reformulating the semiclassical equations of motion for carriers in disordered conductors by taking the quantum mechanical density matrix as the starting point. We focus on the case when only external electric fields are present, without magnetic fields, and the disorder potential is spin-independent. The density matrix approach allows averaging over impurity configurations, and the trace of the velocity operator with the disorder-averaged density matrix can be reinterpreted as the semiclassical velocity weighted by the Boltzmann distribution function. Through this rationale the well-known intrinsic group and anomalous velocities are trivially recovered, while we demonstrate the existence of an \textit{extrinsic interband velocity}, namely a disorder correction to the semiclassical velocity of Bloch electrons, mediated by the interband matrix elements of the Berry connection. A similar correction is present in the non-equilibrium expectation value of the spin operator, contributing to spin-orbit torques. To obtain agreement with diagrammatic approaches the scattering term in the Boltzmann equation must be corrected to first order in the applied electric field, and the Boltzmann equation itself must be solved up to sub-leading order in the disorder potential. Our prescription ensures all vertex corrections present in diagrammatic treatments are taken into account, and to illustrate this we discuss model cases in topological insulators, including the anomalous Hall effect as well as spin-orbit torques. 
\end{abstract}
\date{\today}

\maketitle

\section{Introduction}


Carrier transport in extended conductors is an inherently semiclassical phenomenon, requiring an effective single-particle description as well as averaging over real space and momentum space degrees of freedom. The semiclassical model\cite{Sundaram1999}, which takes these ingredients as natural building blocks, has been a staple of transport theory for the best part of a century\cite{Ashcroft76}. In recent years, aside from its well-established application to the anomalous Hall effect and dynamics in magnetic systems
\cite{Jungwirth2002,Matsumoto2011,ChengRan2016,GaoYang2019,Zhang2019},
it has frequently been used to describe transport in electric and magnetic fields in systems with non-trivial topological textures
\cite{SHINDOU2005,Dumitrescu2012,Son2013,SongJustin2015,YangShengyuan2015,Lensky2015,ChangMingChe2015,Gorbar2018,Misaki2018,Araki2018,Alexandradinata2018,DasKamal2019,Nandy2019,Hou2019,ZhuPenghao2021,Yokoyama2021,WangZhi2021,DasKamal2021,Chaudhary2021,Bhowal2021},
including non-linear electromagnetic responses \cite{Moore2010, Sodemann2015, Morimoto2016, ZhenChuanchang2019, Golub2020} and has recently found substantial applications in computational approaches to non-equilibrium physics
\cite{Wang2006,Wang2007,Gradhand_2012,HeYan2012,ChenY2013,RaffaelloBianco2014,ChenHua2014,Olsen2015,WanxiangFeng2016,XinDai2017,Martiny2019,Wuttke2019,DuShiqiao2020,KTLaw2020,HeWen2021}. In addition to its profound physical insight, broad applicability, and relative simplicity, the semiclassical method naturally accounts for topological effects, and enables a clear identification of Fermi surface and Fermi sea contributions to transport\cite{Chang_2008,Culcer2017,Stedman_2019,Stedman2020}.

The central idea of the semiclassical model is the separation between the dynamics of individual carriers and the carrier distribution. Carrier dynamics between collisions are described by the semiclassical equations of motion, which do not incorporate disorder, while collisions are taken into account through the Boltzmann equation, and affect solely the distribution function, inducing changes in the occupation of quantum states\cite{Ashcroft76}. The semiclassical velocity, originally assumed to be simply the band group velocity, is now known to incorporate a transversal \textit{anomalous} component linear in the driving electric field and proportional to the Berry curvature ${\bm \Omega}_m$ of a given band $m$. \cite{Sundaram1999, XiaoDi2010} Although written in terms of the curvature for a single band this \textit{anomalous velocity} includes inter-band coherence effects, and is associated with band mixing by an electric field.\cite{Culcer2017} The anomalous velocity lies at the heart of the quantum Hall effect and of the intrinsic contribution to the anomalous Hall effect, together with its quantized counterpart. In recent years, however, it has been realised that disorder itself leads to band-mixing effects which are not captured by the Boltzmann equation, and are challenging to include in the wave packet description, since averaging over disorder configurations cannot be done at the level of the wave function. It is well established that a naive application of the semiclassical model to the anomalous and spin-Hall effects in disordered systems makes inaccurate predictions\cite{Inoue_RashbaSHE_Vertex_PRB04, AHE_vertex_PRL_2006}. Indeed, the role of disorder in the anomalous Hall effect,\cite{Karplus_54, Luttinger_AHE_PR58, Smit_SS_58, Berger_SJ_PRB1970, NozLew_SSSJ_JP73} and its relationship to semiclassical dynamics, remains an intensely researched topic\cite{Wang2007, Kovalev2010,YangShengyuan2011,RaffaelloBianco2014,Ado_2015,Ado2016,Ado2017-2,Rauch2018,Keser2019,ShenJianlei2020}. 

Nevertheless, since transport is fundamentally semiclassical, all transport-related quantities must be expressible in semiclassical terms. The assumptions behind wave packet dynamics and diagrammatic approaches are the same: external fields are treated classically and are assumed to be slowly-varying in space, a separation is made between scattering processes and the dynamics between scattering events, and the calculation is performed in the regime $\epsilon_F \tau/\hbar \gg 1$, where $\epsilon_F$ is the Fermi energy and $\tau$ the momentum relaxation time. Recent work has investigated strategies for incorporating the findings of diagrammatic linear response theory into semiclassical dynamics. Xiao and Niu showed that agreement is obtained with diagrammatic approaches if all semiclassical quantities are dressed by disorder, the cost being that the introduction of a disorder-dressed Berry curvature \cite{XiaoCong_2017}. Sinitsyn \textit{et al} introduced a spin-dependent coordinate shift into the position operator and identified a \textit{side-jump velocity}.\cite{Sinitsyn2005, Sinitsyn_2007, Sinitsyn_2007JPCM} These quantities, however, are difficult to work with, since a spin-dependent position operator introduces complications of its own. 

In light of the above, in this paper we formulate the  semiclassical equations of motion for Bloch electrons so as to include disorder using the density matrix formalism, \cite{Culcer2017} in the process making a connection to Green's functions approaches. A strong motivation for our work is the recent surge in applications of the semiclassical model in computational studies of transport in topological materials. 
\cite{Wang2006,Wang2007,Gradhand_2012,HeYan2012,ChenY2013,RaffaelloBianco2014,ChenHua2014,Olsen2015,WanxiangFeng2016,XinDai2017,Martiny2019,Wuttke2019,DuShiqiao2020,KTLaw2020,HeWen2021}
Our primary aim is to provide a straightforward method to incorporate disorder into such computational strategies once a model of disorder is chosen. Taking the density matrix as the starting point\cite{Vasko2005} allows one to average over disorder configurations, something that cannot be done using a wave function. We demonstrate that disorder affects not only the state occupation but also the semiclassical equations of motion, and that it generates a correction to the velocity that accounts for band mixing mediated by the Berry connection and disorder. This approach enables one to distinguish disorder effects on the distribution function from disorder effects on carrier dynamics, yet entails a change in one's point of view so as to regard the semiclassical equations as describing carrier propagation averaged over many disorder scattering events. The carrier undergoes transitions between bands as it scatters, and its trajectory can be determined by averaging over impurity configurations. Whereas the equation of motion for the wave vector ${\bm k}$ follows trivially from the time derivative of the momentum operator, our central result is the revised semiclassical equation of motion for the position of a carrier in band $m$, with dispersion $\epsilon_m$, propagating under the action of an electric field ${\bm E}$ in the presence of disorder
\begin{equation}
\arraycolsep 0.3 ex
\begin{array}{rl}
\displaystyle \dot{\bm r}_m = & \displaystyle \frac{1}{\hbar}\pd{\epsilon_m}{\bm k} +\frac{e}{\hbar} \, {\bm E} \times {\bm \Omega}_m + {\bm \beta}^m_{\bm k}.
\end{array}
\end{equation}
We identify a new contribution to the velocity, which we term the \textit{extrinsic interband velocity} ${\bm \beta}^m_{\bm k}$, defined as:
\begin{equation}
{\bm \beta}^m_{\bm k} =
\frac{1}{2\pi\hbar} 
\int^{\infty}_{-\infty} d\epsilon 
\langle [U,G^A_0(\epsilon)[U,\bm{\mathcal{R}}']G^R_0(\epsilon)]\rangle^{mm}_{\bm k}, 
\end{equation}
where $\bm{\mathcal{R}}'$ represents the inter-band Berry connection with matrix elements $\bm{\mathcal{R}}^{'mn}_{{\bm k}} = i\langle u^{m}_{\bm k}|\nabla_{\bm k} u^{n}_{\bm k}\rangle_{m \ne n}$, and $| u^{m}_{\bm k}\rangle$ is the lattice-periodic part of the Bloch wave function. The extrinsic interband velocity ${\bm \beta}^m_{\bm k}$ is proportional to the disorder strength, which is typically quantified by the impurity density $n_i$, scattering potential strength $u^2_0$, or alternatively $1/\tau$, where $\tau$ is the characteristic scattering time. Formally, ${\bm \beta}^m_{\bm k}$ is similar to the customary scattering term in the Born approximation, except the distribution function is replaced by the band off-diagonal elements of the Berry connection. Since only the band off-diagonal elements of the Berry connection appear ${\bm \beta}^m_{\bm k}$ is by construction gauge covariant. In general ${\bm \beta}^m_{\bm k}$, being independent of applied fields, can be thought of as a disorder-dependent correction to the semiclassical band velocity, or a random inter-band walk on the Fermi surface. We find that ${\bm \beta}^m_{\bm k}$ is nonzero in systems in which time reversal symmetry is broken by e.g. a magnetization. Whereas ${\bm \beta}^m_{\bm k}$ is similar to the \textit{side jump} as defined in Ref.~[\onlinecite{Sinitsyn2006}], unlike Ref.~\onlinecite{Sinitsyn2006}, the present formalism does not employ coordinate shifts, so that the formal and physical position operators coincide. Furthermore, unlike Ref.~[\onlinecite{XiaoCong_2017}], the Berry curvature is the same as in the clean system, rather than being dressed by disorder. Importantly, we show that the scattering term in the Boltzmann equation, needed to determine the effective distribution function, acquires a correction to first order in the electric field, which is equivalent to a gradient expansion in the electrostatic potential. In addition, the Boltzmann equation needs to be solved up to the sub-leading order in the impurity strength, in order to incorporate processes customarily termed skew scattering and side jump.\cite{Smit_SS_58, Berger_SJ_PRB1970, Nagaosa2010} The method we present here also enables us to calculate spin densities using the semiclassical model and obtain accurate results for spin-orbit torques. In order to accomplish this the bare spin expectation value needs to be supplemented with an electric field contribution, and we find an analogous quantity to ${\bm \beta}^m_{\bm k}$ in the spin expectation value. 

More generally, we present a prescription for mapping steady-state expectation values onto the semiclassical model by expressing traces purely in terms of the band diagonal elements of the density matrix. Since in linear response theory all expectation values are traced back to the equilibrium density matrix, which is band-diagonal, they can all be recast in terms of semiclassical quantities. The band-diagonal elements of the density matrix represent the Boltzmann distribution, which can be evaluated from the much simpler Boltzmann equation. In fact we will argue briefly in the latter part of this work that linear response theories can be thought of as a family tree with its roots in the quantum Liouville equation: the Kubo approach is the integral formulation, the quantum kinetic, or quantum Boltzmann, approach is the integro-differential formulation, while the semiclassical model is an offshoot of the latter, which arises as a result of an additional separation between the carrier dynamics and distribution. Sharing a common origin, these methods yield equivalent results, and, in particular, vertex corrections present in diagrammatic approaches have straightforward equivalents in the semiclassical language. The blueprint presented in this work can be used in the future to incorporate electron-electron interactions into computational approaches in a mean field picture.

The outline of this paper is as follows. In Sec. \eqref{Sec:ModelHamiltonian} we introduce the Hamiltonian and the model of disorder. In Sec.\eqref{Sec:GeneralTheoryDM} we review linear response theory based on the density matrix and introduce the electric-field correction to the collision term. Next in Sec.\eqref{Sec:TheoryHallEffectSpinDensity} we outline the general methodology for deriving the semiclassical equations of motion from the quantum kinetic equation, and discuss also disorder effects on spin expectation values. In Sec. \eqref{Sec:FamilyTreeLRT} we analysed the relation among different linear response methodologies commonly used to calculate transport coefficients. In Sec.\eqref{Sec:Applications} we discuss at length two model examples, the anomalous Hall effect and spin-orbit torques in magnetic topological insulators. We end with a summary and conclusions. 

\section{Model Hamiltonian}
\label{Sec:ModelHamiltonian}

We consider a Hamiltonian of the form: 
\begin{equation}
\label{ }
H=H_0 + V(\bm r) + U(\bm r),
\end{equation}
with $H_0$ the low energy effective band Hamiltonian, in principle assumed to include the Zeeman interaction with an external magnetic field, $V(\bm r)$ is the electrostatic potential, and $U=U(\bm r)$ represents the disorder scattering potential. We emphasize the Hamiltonian is always Hermitian. Non-Hermitian systems were considered in Ref.~[\onlinecite{Silberstein2020}].

We work in the crystal momentum representation $|m,\bm k\rangle=e^{i\bm k \cdot \bm r}|u^{m}_{\bm k}\rangle$. The matrix elements of a scalar disorder potential $U(\bm r)$ are given by the equation 
\begin{equation}
\label{Eq:ScatteringAmplitud}
U^{mm'}_{\bm k\bm k'}= \langle u^{m}_{\bm k} |u^{m'}_{\bm k'} \rangle \mathcal{U}_{\bm q},
\end{equation}
where we have defined the Fourier transform of the spatial function in $d$ dimensions 
\begin{equation}
\label{Eq:FourierTransformPotential}
\mathcal{U}_{\bm q}=
\int d^d\bm r U(\bm r )e^{-i\bm q \cdot \bm r},
\end{equation}
with $\bm q=\bm k-\bm k'$. The impurity average is defined by 
\begin{equation}
\label{ }
\langle U^{mm'}_{\bm k\bm k'}U^{nn'}_{\bm k'\bm k} \rangle= \langle u^{m}_{\bm k} |u^{m'}_{\bm k'}\rangle\langle u^{n'}_{\bm k'} |u^{n}_{\bm k} \rangle 
\langle \mathcal{U}_{\bm q} \mathcal{U}_{-\bm q}\rangle,
\end{equation}
with 
\begin{equation}
\label{ }
\langle \mathcal{U}_{\bm q} \mathcal{U}_{-\bm q}\rangle = \int\int d^d\bm r d^d\bm r' \langle U(\bm r )U(\bm r' )\rangle e^{-i\bm q \cdot (\bm r-\bm r')},
\end{equation}
where $\langle \cdots \rangle$ refers to an average over impurity configurations. For concreteness we will use a model of disorder whose spatial correlations function is defined as:
\begin{align}
\label{}
\langle U(\bm r)\rangle &=0 \\
\langle U(\bm r)U(\bm r')\rangle &=u^2_{0}\delta(\bm r-\bm r') 
\end{align} 

Then, it follows that 
\begin{equation}
\label{}
\langle U^{mm'}_{\bm k\bm k'}U^{nn'}_{\bm k'\bm k} \rangle
=u^2_0\langle u^{m}_{\bm k} |u^{m'}_{\bm k'}\rangle\langle u^{n'}_{\bm k'} |u^{n}_{\bm k} \rangle,
\end{equation}
where $u^2_0$ is a parameter that takes into account the strength of the disorder potential. 

\section{Quantum Kinetic Equation}
\label{Sec:GeneralTheoryDM}

In this section we give a brief presentation of the quantum kinetic equation in a somewhat different language than that used in Ref.~[\onlinecite{Culcer2017}]. We note that similar density-matrix based approaches have been used recently to describe carrier dynamics in the semiclassical regime\cite{Stedman_2019,Stedman2020}. 
The starting point is the quantum Liouville equation for the single particle density operator $\rho$, namely: 
\begin{equation}\label{Liouville}
\frac{\partial \rho}{\partial t} + \frac{i}{\hbar}[H,\rho] = 0.
\end{equation}
For the sake of convenience, we introduce at this stage the free retarded Green function
\begin{equation}
\label{}
G^R_0(t)=-i\theta(t)e^{-itH_0/\hbar},
\end{equation}
In the frequency domain 
\begin{align}
\label{}
G^R_0(\epsilon)
&=-\frac{i}{\hbar}\int^{\infty}_{0} dt e^{-iH_0t/\hbar} e^{i\epsilon t/\hbar} e^{-\eta t},
\end{align}
where we introduced the factor $e^{-\eta t}$ to ensure convergence. The advanced Green function follows by Hermitian conjugation. 

\subsection{Kinetic equation in equilibrium}

For the sake of simplicity, let us for now ignore the effect of the driving electric field in the kinetic equation. Using a decomposition of the density matrix as $\rho = \langle\rho\rangle + g_0$ in the quantum Liouville equation\cite{Culcer2017}, we get for the disorder averaged part in equilibrium:
\begin{equation}
\label{Eq:KineticEquationOne}
\frac{\partial \langle\rho \rangle}{\partial t}
+\frac{i}{\hbar}[H_0,\langle\rho \rangle]+\frac{i}{\hbar}\langle [U,g_0]\rangle=0.
\end{equation}
while for $g_0$ we get the equation:
\begin{equation}
\label{Eq:KineticEquationTwo}
\frac{\partial g_0}{\partial t}+
\frac{i}{\hbar}[H_0,g_0]+\frac{i}{\hbar}[U,g_0]-\frac{i}{\hbar}\langle [U,g_0]\rangle=-\frac{i}{\hbar} [U,\langle\rho \rangle].
\end{equation}
In order to solve the kinetic equation for $\langle\rho \rangle$, we first solve Eq.\eqref{Eq:KineticEquationTwo} for $g_0$ and then we use it in Eq.\eqref{Eq:KineticEquationOne}. In the first Born approximation \cite{Culcer2017} we neglect the last two terms on the left hand side of eq.\eqref{Eq:KineticEquationTwo}. We are left with
\begin{equation}
\label{}
\frac{\partial g_0}{\partial t}+\frac{i}{\hbar}[H_0,g_0]=-\frac{i}{\hbar} [U,\langle\rho \rangle].
\end{equation}
Solving for $g_0$
\begin{align}
\label{}
g_0 &=-\frac{i}{\hbar}\int^{\infty}_{0}dt'  [e^{-iH_0t'/\hbar}U e^{iH_0t'/\hbar} ,\langle\rho (t) \rangle].
\end{align}
In terms of Green's functions $g_0$ can be expressed as
\begin{align}
\label{Eq:solution-g}
g_0
&=\frac{1}{2\pi i}\int^{\infty}_{0}d\epsilon [G^{R}_{0}(\epsilon)U G^{A}_{0}(\epsilon) ,\langle\rho (t) \rangle]. 
\end{align}
This solution is substituted into Eq. \eqref{Eq:KineticEquationOne}. We arrive at the equation
\begin{equation}
\label{}
\frac{\partial \langle\rho \rangle}{\partial t}+\frac{i}{\hbar}[H_0,\langle\rho \rangle]+J(\langle\rho \rangle)=0,
\end{equation}
with the collision integral $J(\langle\rho \rangle)$ defined as:
\begin{equation}
\label{ }
J(\langle\rho \rangle)=\frac{i}{\hbar}\langle [U,g_0]\rangle. 
\end{equation}

\subsection{Adding an electric field}

Let us now consider the effect of the driving electrostatic potential up to linear order. For simplicity we take this potential to have the form $V(\bm r)=e\bm E\cdot\bm r$, implying a uniform electric field, which corresponds to the overwhelming majority of experimental setups. The case of inhomogeneous systems, including systems in inhomogeneous electric fields, entails additional subtleties which we postpone for later consideration\cite{Marrazzo2017,Lapa2019,Matisse2020,Kozii2021}. 
Adding an electric field to the Hamiltonian implies a correction to the function $g$, which can then be written as $g = g_0 + g_E$, where $g_0$ was found in the previous section, and
\begin{equation}
\frac{\partial g_E}{\partial t}+\frac{i}{\hbar}[H_0,g_E] = -\frac{i}{\hbar}[V,g_0].
\end{equation}
The notation $g_E$ reflects the fact that eventually it is the electric field that appears in the final expressions, rather than the electrostatic potential. For $g_E$ we find explicitly
\begin{align}
\label{Eq:ModifiedSolutionToCollisionIntegral}
g_E & = -\frac{i}{\hbar}
\int^{\infty}_{0}dt'' e^{-iH_0t''/\hbar} [V , g_{0}(t-t'')] e^{iH_0t''/\hbar}.
\end{align}
The function $g_E$ is off-diagonal in the momentum as well as in the band index. We solve Eq.\eqref{Eq:ModifiedSolutionToCollisionIntegral} by introducing Markovian approximation which reads $g_0(t-t'')\approx g_0(t)$, whereupon in the frequency domain we obtain
\begin{align}
\label{Eq:FrequencyDomain_g(E)}
g_E & = \frac{1}{2\pi i}\int^{\infty}_{0}d\epsilon G^{R}_{0}(\epsilon)[V, g_0 ] G^{A}_{0}(\epsilon),
\end{align}
and in the commutator, we should use Eq.\eqref{Eq:solution-g} as a functional of the equilibrium distribution function $f_0(\epsilon)$ to fulfill linear response. 

The kinetic equation for the disorder averaged density  matrix $\langle\rho \rangle$ is now modified to
\begin{equation}
\label{Eq:KineticEquationThree}
\frac{\partial \langle\rho \rangle}{\partial t} + \frac{i}{\hbar}[H_0,\langle\rho \rangle] + J_{0}(\langle\rho \rangle) = -\frac{i}{\hbar}[V,\langle\rho \rangle] - J_E(\langle\rho \rangle),
\end{equation}
with the collision integral $J_E(\langle\rho \rangle)$ defined as:
\begin{align}
\label{Eq:modifiedCollisionIntegral}
J_E(\langle\rho \rangle) &= \frac{i}{\hbar}\langle [U,g_E] \rangle. 
\end{align}
As we show below, the electric field correction to the collision integral in Eq. \eqref{Eq:modifiedCollisionIntegral} with the off-diagonal density function as given in 
Eq. \eqref{Eq:FrequencyDomain_g(E)} will provide results in agreement with previous calculations based on diagrammatic perturbation theory\cite{Ado_2015,Ndiaye_2017}. We consider such an agreement as a positive test of the Markovian approximation. We note also that Eq.~\eqref{Eq:FrequencyDomain_g(E)} was used in a different but equivalent  form in a previous paper \cite{Culcer2010} in order to calculate side jump effects in a system with extrinsic spin-orbit coupling. In this paper, we will focus on systems with intrinsic spin-obit coupling. 

\subsection{Kinetic equation and linear response}

When Eq.~\eqref{Eq:KineticEquationThree} is expressed in the crystal momentum representation we obtain the quantum kinetic equation\cite{Sekine2017}
\begin{equation}\label{QKE}
    \frac{\partial f_{\bm k}}{\partial t} + \frac{i}{\hbar} \, [H_{0{\bm k}}, f_{\bm k}] + J_{0}(f_{\bm k}) = \frac{e{\bm E}}{\hbar}\cdot\frac{Df_{\bm k}}{D{\bm k}} - J_E(f_{\bm k}).
\end{equation}
We have written the matrix elements of $\bkt{\rho}$ in this representation as $f_{\bm k}$. We refer to $f_{\bm k}$ henceforth as the density matrix, noting that it has matrix elements connecting different bands, although the band index $n$ has not been written explicitly. The covariant derivative $\frac{Df_{\bm k}}{D{\bm k}} = \frac{\partial f_{\bm k}}{\partial{\bm k}} -
i[\bm{\mathcal{R}}_{\bm k}, f_{\bm k}]$.

To solve Eq.~\ref{QKE}, the density matrix is separated into a band diagonal and a band off-diagonal part, namely, we write $f_{\bm k} = n_{{\bm k}} + S_{{\bm k}}$. The band diagonal term $n_{{\bm k}}$ represents the fraction of carriers in a specific band and is essentially the solution of the ordinary Boltzmann equation, while $S_{{\bm k}}$ contains the effect of inter-band coherence, or band mixing. All our effort in recovering the semiclassical theory consists of eliminating $S_{{\bm k}}$. The effective Boltzmann equation that we shall derive is simply what is obtained for $n_{{\bm k}}$ once all references to $S_{{\bm k}}$ have been eliminated. Fortunately, as we recapitulate below, the solution for $S_{{\bm k}}$ in an electric field is relatively simple, making it straightforward to express expectation values in terms of $n_{{\bm k}}$ alone. 

The equilibrium density matrix is band diagonal, its elements represented by the Fermi-Dirac distribution for each band $n_{FD}(\epsilon^m_{\bm k})$. In an electric field one may expand to linear order $f_{\bm k}^{mn} = n_{FD}(\epsilon^m_{\bm k})\delta_{mn} + f_{E{\bm k}}^{mn}$, with corresponding expressions for $n_{E{\bm k}}$ and $S_{E{\bm k}}$. The kinetic equation is split into two coupled equations for $n_{E{\bm k}}$ and $S_{E{\bm k}}$, whose solution, based on an expansion in the small parameter $\hbar/(\epsilon_F\tau)$, is explained in detail in Ref.~[\onlinecite{Culcer2017}]. It was shown that $n_{E{\bm k}}$ starts at order $-1$ in this small parameter, since it is proportional to the scattering time $\tau$, while $S_{E{\bm k}}$ starts at order $0$. Consequently, the sub-leading correction to $n_{E{\bm k}}$, referred to as $n_{E{\bm k}}^{(0)}$, is also required.

To leading order in $\hbar/(\epsilon_F\tau)$, the diagonal part reads:   
\begin{equation}\label{Boltzdiag}
[J_{0}(n_E^{(-1)})]^m_{{\bm k}} = \frac{e\bm E}{\hbar}\cdot \frac{\partial n_{FD}(\epsilon^{m}_{\bm k})}{\partial \bm k}, 
\end{equation}
where the Born approximation collision integral is
\begin{align}
\label{Eq:CollisionIntegralEquation}
[J_0(n_E)]^m_{{\bm k}}&=
\frac{2\pi}{\hbar}
\sum_{m',\bm k'}
\langle U^{mm'}_{\bm k \bm k'}U^{m'm}_{\bm k'\bm k}\rangle 
\\ \nonumber
&\times 
\left(
n^{m}_{E\bm k}-n^{m'}_{E\bm k'}
\right)
\delta(\epsilon^{m}_{\bm k}-\epsilon^{m'}_{\bm k'}).
\end{align}

The solution of Eq. \eqref{Boltzdiag} is in general rather complicated\cite{Allen1978}. For a system with isotropic dispersion it reduces to the simple form
\begin{equation}
\label{nDrude}
n_{E{\bm k}}^{m(-1)} =\tau^{m}_p \frac{e\bm E}{\hbar}\cdot 
\frac{\partial \epsilon^{m}_{\bm k}}{\partial \bm k} 
\frac{\partial n_{FD}(\epsilon^{m}_{\bm k})}{\partial \epsilon^{m}_{\bm k}}, 
\end{equation}
where the transport time $\tau^m_p$ is defined as:
\begin{equation}
\label{Eq:transportTime}
\frac{1}{\tau^{m}_p} = \frac{2\pi}{\hbar}\sum_{m',\bm k'}
\langle U^{mm'}_{\bm k \bm k'} U^{m'm}_{\bm k' \bm k}\rangle
[1-\cos(\theta_{\bm k'}-\theta_{\bm k})]\delta(\epsilon^{m}_{\bm k}-\epsilon^{m'}_{\bm k'}).
\end{equation}
The solution for $S^{(0)}_{E\bm k}$ takes the simple form\cite{Culcer2017}
\begin{equation}
\label{Eq:OffDiagonalDM}
S^{(0)mm'}_{E{\bm k}}
=\frac{\hbar (D + D')^{mm'}_{E\bm k}}{i(\epsilon^{m}_{\bm k}-\epsilon^{m'}_{\bm k}-i\eta)}
\end{equation}
with the intrinsic and anomalous driving terms, \cite{Culcer2017} 
\begin{widetext}
\begin{align}
\label{Eq:IntrinsicDriving}
D^{mm'}_{E\bm k} & = \frac{ie}{\hbar}\bm E\cdot 
\bm{\mathcal{R}}^{mm'}_{\bm k}
[n_{FD}(\epsilon^m_{\bm k}) - n_{FD}(\epsilon^{m'}_{\bm k})] \\
\label{Eq:AnomalousDriving}
D'^{mm'}_{E\bm k} & = - \frac{\pi}{\hbar}
\sum_{m'',\bm k'}
\langle U^{mm''}_{\bm k \bm k'}U^{m''m'}_{\bm k'\bm k}\rangle 
 \left\{(n_{E{\bm k}}^{m'} - n_{E{\bm k'}}^{m''})
\delta(\epsilon_{m'\bm k}-\epsilon_{m''{\bm k'}}) 
\right. + \left. (n_{E{\bm k}}^m - n_{E{\bm k'}}^{m''})
\delta(\epsilon_{m''{\bm k'}} - \epsilon_{m{\bm k}})\right\}.
\end{align}
\end{widetext}

Since $S_{E\bm k}$ starts at zeroth order in the parameter $\hbar/(\epsilon_F\tau)$, we also require the sub-leading term $n^{(0)}_{E\bm k}$, which is found from the equation
\begin{equation}
\label{Eq:Subleading}
[J_0(n^{(0)}_{E})]^m_{\bm k} = - [J_{sk}(n^{(-1)}_{E})]^{m}_{\bm k} - [J_{E}(n_{FD})]^m_{\bm k},
\end{equation}
where the right hand side acts as the driving term, whose constituents will be explained shortly. Solving this equation will yield two different contributions to the sub-leading diagonal density matrix $n^{(0)}_{E{\bm k}}$, which we write as $n^{(0)}_{E{\bm k}}=n^{(\rm sk)}_{E{\bm k}}+n^{(\rm sj)}_{E{\bm k}}$. Although both $n^{(\rm sj)}_{E{\bm k}}$ and $n^{(\rm sk)}_{E{\bm k}}$ are of zeroth order in $\hbar/(\epsilon_F\tau)$, they are parametrically different with respect to magnetisation and Fermi energy, as we will see later on. We can solve for these two terms separately as follows.

The contribution $n^{(\rm sk)}_{E{\bm k}}$ stems from $D'$ and is associated with skew scattering in the semiclassical theory. It is solved in an analogous manner to Eq.~\eqref{Boltzdiag}, namely, the driving term is found by substituting Eq.~\eqref{Eq:OffDiagonalDM} into a collision integral of the form of Eq.~\eqref{Eq:CollisionIntegralEquation}, obtaining
\begin{equation}
\label{Eq:balanceEquationSkew}
J_0[n^{(\rm sk)}_{E}] = - [J_{sk}(n^{(-1)}_{E})]^{m}_{\bm k},
\end{equation}
which can be solved for $n^{(\rm sk)}_{E{\bm k}}$ using the standard techniques of Boltzmann theory\cite{Allen1978}. The driving term in this equation can be written explicitly as a function of the leading-order density matrix $n^{(-1)}_{E{\bm k}}$ as
\begin{widetext}
\begin{align}
\label{}
\nonumber
[J_{sk}(n^{(-1)}_{E})]^{m}_{\bm k}
=
\frac{2\pi^2}{\hbar} 
\sum_{m'm''n{\bm k}'{\bm k}''}
& {\rm Im}\left[\frac{ \langle U^{mm''}_{\bm k \bm k'}U^{m'm}_{\bm k'\bm k}\rangle 
\langle U^{m''n}_{\bm k' \bm k''}U^{nm'}_{\bm k''\bm k'}\rangle }
{(\epsilon^{m''}_{\bm k'}-\epsilon^{m'}_{\bm k'})}\right] \\
& \left\{
(n_{E{\bm k'}}^{m'(-1)} - n_{E{\bm k''}}^{n(-1)})
\delta(\epsilon^{m'}_{\bm k'}-\epsilon^{n}_{\bm k''}) 
+
(n_{E{\bm k'}}^{m''(-1)} - n_{E{\bm k''}}^{n(-1)} )
\delta(\epsilon^{n}_{\bm k''}-\epsilon^{m''}_{\bm k'})
\right\} 
\delta(\epsilon^{m''}_{\bm k'}-\epsilon^{m}_{\bm k}) \\ \nonumber
-
\frac{2\pi^2}{\hbar} 
\sum_{m'm''n{\bm k}'{\bm k}''}
&{\rm Im}\left[
\frac{ \langle U^{m''m'}_{\bm k \bm k'}U^{m'm}_{\bm k'\bm k}\rangle
\langle U^{mn}_{\bm k \bm k''}U^{nm''}_{\bm k''\bm k}\rangle }
{(\epsilon^{m}_{\bm k}-\epsilon^{m''}_{\bm k})}\right]
\\ \nonumber
& \left\{
(n_{E{\bm k}}^{m''(-1)} - n_{E{\bm k''}}^{n(-1)})
\delta(\epsilon^{m''}_{\bm k}-\epsilon^{n}_{\bm k''})  
+
(n_{E{\bm k}}^{m(-1)} - n_{E{\bm k''}}^{n(-1)})
\delta(\epsilon^{n}_{\bm k''}-\epsilon^{m}_{\bm k})
\right\} 
\delta(\epsilon^{m'}_{\bm k'}-\epsilon^{m''}_{\bm k}),
\end{align}
\end{widetext}
and we recall that $n_{E{\bm k}}^{(-1)}$ was found in Eq.~\eqref{nDrude}.

The second contribution to the driving term in Eq.~\eqref{Eq:Subleading} is due to the electric field correction of the collision integral $J_E$ acting on the equilibrium distribution function. Since this contribution is associated with side jump scattering in the semiclassical theory, we will refer to it as $n^{(\rm sj)}_{E{\bm k}}$. To determine $n^{(\rm sj)}_{E{\bm k}}$ we need to solve the equation
\begin{equation}
\label{Eq:balanceEquationSideJump}
J_0[n^{(\rm sj)}_{E}]^m_{\bm k} = -[J_{E}(n_{FD})]^{m}_{\bm k}.
\end{equation}
In the crystal momentum representation the electric field correction to the collision integral takes the form
\begin{widetext}
\begin{align}
\nonumber
\label{Eq:ElectricFieldCorrectionCollisionIntegral}
[J_E(n_{FD})]^{m}_{\bm k}
&=
\frac{2\pi}{\hbar}
\frac{\partial n_{FD}(\epsilon^{m}_{\bm k}) }{\partial \epsilon^{m}_{\bm k}}
e\bm E \cdot
\sum_{\bm k'}
\langle U^{mm}_{\bm k \bm k'}U^{mm}_{\bm k'\bm k}\rangle
\left[
\bm{\mathcal{R}}^{mm}_{\bm k'} - \bm{\mathcal{R}}^{mm}_{\bm k}
\right]
\delta(\epsilon^{m}_{\bm k'}-\epsilon^{m}_{\bm k})\\
&+
\frac{2\pi }{\hbar}
\frac{\partial n_{FD}(\epsilon^{m}_{\bm k})}{\partial \epsilon^{m}_{\bm k}} 
e\bm E\cdot
\sum_{m'\bm k'}
{\rm Im}\left\{\left\langle 
\left[
\left(\nabla_{\bm k}+\nabla_{ \bm k'}\right)U^{mm'}_{\bm k\bm k'} \right] 
U^{m'm}_{\bm k' \bm k} \right\rangle \right\} 
\delta(\epsilon^{m'}_{\bm k'}-\epsilon^{m}_{\bm k}),
\end{align}
\end{widetext}
where the derivatives act only on $U^{mm'}_{\bm k\bm k'}$. This equation should be compared with the side jump velocity calculated from a coordinate shift introduced in Ref.[\onlinecite{Sinitsyn2006}]. The balance between two collision integrals, as stated in Eq. \eqref{Eq:balanceEquationSideJump} provides the necessary information to calculate a new subleading density function $n^{(\rm sj)}_{E}$ that in the semiclassical language \cite{Sinitsyn2006_II} is interpreted as an anomalous distribution due to coordinate shift of the scattered particle after many collisions.   

\section{Recovering the semiclassical theory}
\label{Sec:TheoryHallEffectSpinDensity}

In this section we decompose the kinetic equation into a part representing carrier dynamics and a part representing the distribution, which is found from a modified Boltzmann equation. Since the equation of motion for the carrier wave vector, yielding $\hbar\dot{\bm k} = -e{\bm E}$, follows immediately from the operator commutator $[{\bm p}, V({\bm r})]$, the bulk of our effort is devoted to finding the disorder-averaged velocity, which will yield the time evolution of the carrier position $\dot{\bm r}_n$. The prescription for recovering the semiclassical theory from the quantum kinetic equation proceeds as follows: 
\begin{itemize}
    \item Determine the velocity expectation value as the operator trace Tr $(\dot{\bm r}f)$, where $\bm{\dot{r}} = \frac{i}{\hbar}[H, \bm r]$ represents the matrix elements of the velocity operator. In the crystal momentum representation these are given by the covariant derivative $\dot{\bm r} = \frac{1}{\hbar}\frac{DH}{D{\bm k}}$. 
    \item Reduce the trace to a form in which \textit{only} band-diagonal elements of the density matrix appear. These will contain either the equilibrium Fermi-Dirac distribution $n_{0{\bm k}}$, or the correction to the band-diagonal part $n_{E{\bm k}}$, which we recall has three constituents: $n_{E{\bm k}} = n_{E{\bm k}}^{(-1)} + n_{E{\bm k}}^{(sk)} + n_{E{\bm k}}^{(sj)}$.
    \item The result follows a natural separation into a contribution associated with the equation of motion $\dot{\bm r}_n$ and one associated with the Boltzmann equation.
    \item For the spin density, we follow similar steps, namely, we take the trace of the spin operator in the Bloch basis with the averaged density matrix. We will also find an extrinsic spin matrix element that accounts for spin rotations during scattering events. 
\end{itemize}

The Hamiltonian is $H = H_0 + V({\bm r}) + U({\bm r})$, and since the last two terms commute with the position operator they do not contribute to the velocity operator. The band Hamiltonian yields
\begin{widetext}
\begin{align}
{\rm Tr}(\dot{\bm r} f) \rightarrow {\rm Tr} \left\{ \frac{i}{\hbar}[H_0, \bm r] f \right\} 
&=
\frac{1}{\hbar}
\sum_{m',m,\bm k}
\left[\frac{\partial \epsilon^{m'}_{\bm k}}{\partial \bm k} \delta_{m,m'}  
+ i(\epsilon^{m}_{\bm k}-\epsilon^{m'}_{\bm k}) \bm{\mathcal{R}}^{mm'}_{\bm k}
\right]f^{m'm}_{\bm k} ,
\\
\label{ }
&=
\sum_{m,\bm k} \bm v^{m}_{\bm k} [n_{E{\bm k}}^{(-1)m}
+ n^{(0)mm}_{\bm E\bm k}] + \frac{i}{\hbar}
\sum_{m',m,\bm k} (\epsilon^{m}_{\bm k} - \epsilon^{m'}_{\bm k}) \bm{\mathcal{R}}^{mm'}_{\bm k} S^{(0)m'm}_{E{\bm k}}.
\end{align}
\end{widetext}
The Berry connection $\bm{\mathcal{R}}^{mm'}_{\bm k}=i\langle u^{m}_{\bm k}|\nabla_{\bm k} u^{m'}_{\bm k}\rangle$. The first term gives the usual group velocity $\bm v^{m}_{\bm k}=\nabla_{\bm k}\epsilon^{m}_{\bm k}/\hbar$ which is diagonal, while the second term gives a contribution due to band mixing and is purely off diagonal. We will concentrate on the second factor or band mixing velocity. The off-diagonal density matrix is composed of two terms: an intrinsic one and an extrinsic one. Let us first consider the intrinsic one. It is 
\begin{align}
\label{Eq:bandmixingvelocityInt}
{\rm Tr}\{\dot{\bm r} f \}^{\rm int}
&=
-\sum_{m',m,\bm k}\bm{\mathcal{R}}^{mm'}_{\bm k}D^{m'm}_{E \bm k}. 
\end{align}
After replacing the driving term $D^{m'm}_{E \bm k}$ by exchanging $m \rightarrow m' $ in the first term, and summing over intermediate states, the intrinsic contribution can be written as the average of the transverse velocity
\begin{align}
\label{ }
{\rm Tr}\{\dot{\bm r} f \}^{\rm int} &=
\frac{e}{\hbar} \sum_{m,\bm k} \bm E\times {\bm\Omega^{m}_{\bm k}} n_{FD}(\epsilon^{m}_{\bm k})
\end{align}
with the Berry curvature 
\begin{equation}
\label{ }
\Omega^{m}_{k,z}
=
i\left[
\left\langle \frac{\partial u^{m}_{\bm k}}{\partial k_x}\Big | \frac{\partial u^{m}_{ \bm k}}{\partial k_y} \right\rangle
-
\left\langle \frac{\partial u^{m}_{\bm k}}{\partial k_y}\Big | \frac{\partial u^{m}_{ \bm k}}{\partial k_x} \right\rangle
\right].  
\end{equation}
The extrinsic contribution reads
\begin{align}
\label{Eq:ExtrinsicAverageVelocity}
{\rm Tr}\{\dot{\bm r} f \}^{\rm ext}
&=
\sum_{n,m,\bm k} \bm{\mathcal{R}}^{nm}_{\bm k}[J(f)]^{mn}_{\bm k}.
\end{align}
After some algebra it can be written as
\begin{align}
\label{betanE}
{\rm Tr}\{\dot{\bm r} f \}^{\rm ext}
=
\langle \bm \beta \rangle
&=
\sum_{m,\bm k} n_{E{\bm k}}^{(-1)m} \bm{\beta}^{m}_{\bm k},
\end{align}
with the function $\bm{\beta}^{m}_{\bm k}$ formally defined as
\begin{align}
\label{ }
\bm{\beta}^{m}_{\bm k}
&=
\frac{1}{2\pi\hbar} 
\int^{\infty}_{-\infty} d\epsilon 
\langle [U,G^A_0(\epsilon)[U,\bm{\mathcal{R}'}]G^R_0(\epsilon)]\rangle^{mm}_{\bm k},
\end{align}
where the prime in $\bm{\mathcal{R}'}$ indicates that only the band off-diagonal matrix elements of the Berry connection enter. In Eq.~\eqref{betanE} we have written directly the electric-field dependent correction to the distribution function, since Eq.~\eqref{Eq:ExtrinsicAverageVelocity} makes it obvious that this contribution vanishes when $f$ is replaced by the equilibrium distribution $n_{FD}$. This is because, for scalar scattering as studied in this work, the equilibrium distribution causes the entire collision integral to vanish. For computational evaluations it will be useful to list the explicit equation for $\bm{\beta}^{m}_{\bm k}$:
\begin{widetext}
\begin{align}
\bm{\beta}^{m}_{\bm k}
&=
\frac{\pi}{\hbar}
\sum_{n,m',\bm k'} 
\Big\{
\left[
\bm{\mathcal{R}}^{'mn}_{\bm k}
\langle
U^{nm'}_{\bm k,\bm k'}
U^{m'm}_{\bm k',\bm k}
\rangle
+
\langle
U^{mm'}_{\bm k,\bm k'}
U^{m'n}_{\bm k',\bm k}
\rangle
\bm{\mathcal{R}}^{'nm}_{\bm k}
\right]
\delta(\epsilon^{m}_{\bm k}-\epsilon^{m'}_{\bm k'}) \\ \nonumber
&-
\left[
\langle U^{mn}_{\bm k,\bm k'} U^{m'm}_{\bm k',\bm k} \rangle \bm{\mathcal{R}}^{'nm'}_{\bm k'} 
+
\langle U^{mm'}_{\bm k,\bm k'} U^{nm}_{\bm k',\bm k} \rangle \bm{\mathcal{R}}^{'m'n}_{\bm k'} 
\right] 
\delta(\epsilon^{n}_{\bm k'}-\epsilon^{m}_{\bm k})\Big\}
\end{align}
\end{widetext}
Note that $\bm{\beta}^{m}_{\bm k}$ is proportional to the disorder strength quantified here by $u^2_0$, making it first order in $\hbar/(\epsilon_F\tau)$. It represents a disorder-dependent correction to the semiclassical band velocity, which is independent of the applied electric field. Physically, $\bm{\beta}^{m}_{\bm k}$ represents the average value of the random changes in the carrier velocity that occur every time the carrier is scattered between bands. Since $\bm{\beta}^{m}_{\bm k}$ has units of velocity and depends on the disorder potential we will refer to it as the \textit{extrinsic inter-band velocity}. Given that $\bm{\beta}^{m}_{\bm k}$ is formally of first order in $\hbar/(\epsilon_F\tau)$, we are only interested in its product with the leading term in the distribution function, $n_{E{\bm k}}^{(-1)}$, so that its overall contribution to the current is formally zeroth order in disorder. Moreover, with $n_{E{\bm k}}^{(-1)}$ representing a Fermi surface contribution, the net effect of $\bm{\beta}^{m}_{\bm k}$ can be thought of as a random inter-band walk on the Fermi surface. Interestingly, ${\bm \beta}_n$ has the same mathematical form as the Born approximation scattering term $J_0$, except the band off-diagonal elements of $\bm{\mathcal{R}}$ appear instead of $n_{\bm k}$. The presence of only the band off-diagonal matrix elements of $\bm{\mathcal{R}}$ ensures $\bm{\beta}^{m}_{\bm k}$ is gauge covariant. In the examples we study below we find that $\bm{\beta}^{m}_{\bm k}$ is nonzero in systems in which time reversal symmetry is broken by e.g. a magnetization. It is similar to the \textit{side jump} appearing in Ref.~[\onlinecite{Sinitsyn2006}], although we stress that our approach makes no reference to any coordinate shifts, and the formal position operator is identical to the physical position operator.

Since all contributions to the current density are now expressed in terms of the distribution function (the equilibrium as well as the leading and sub-leading terms in an electric field), we are able to write the semiclassical equations of motion as
\begin{align}
\label{Eq:SemiclassicalEffectiveVelocity}
\dot{\bm r}_{m}&=\frac{1}{\hbar}\frac{\partial \epsilon^{m}_{\bm k}}{\partial \bm k} - \dot{\bm k}_{m}\times \bm{\Omega}^{m}_{\bm k}+\bm{\beta}^{m}_{\bm k} \\
\hbar \dot{\bm k}_{m} &=-e \bm E.
\end{align}

The distribution function is found from the Boltzmann equation, with the caveat that we require both the leading and subleading order terms in the disorder strength. The procedure is as follows. First the leading-order term in the distribution function $n_{E{\bm k}}^{(-1)}$ is found from
\begin{equation}
\label{Boltzeq}
J_{0}[n_{E}^{(-1)}]^m_{\bm k}
=\frac{e\bm E}{\hbar}\cdot \frac{\partial n_{FD}(\epsilon^{m}_{\bm k})}{\partial \bm k}, 
\end{equation}
while the sub-leading correction $n^{(0)}_{E}$ is given by
\begin{equation}
\label{BoltzSub}
J_{0}[n^{(0)}_{E}]^m_{\bm k} = - [J_{E}(n_{FD})]^m_{\bm k} - J_{sk}[n_{E}^{(-1)}]^m_{\bm k},
\end{equation}
where the left hand side is the quantity to be found, and the right hand side plays the role of a driving term. Finally, we are able to write the full expectation value of the current in terms of semiclassical quantities
\begin{widetext}
\begin{equation}
    \bkt{\bm j} = (-e) \, \sum_{m{\bm k}} \bigg\{ \frac{1}{\hbar}\frac{\partial \epsilon^{m}_{\bm k}}{\partial \bm k} [n^{(-1)m}_{E{\bm k}} + n^{(sk)m}_{E{\bm k}} + n^{(sj)m}_{E{\bm k}}] + \bigg(\frac{e\bm E}{\hbar} \times \bm{\Omega}^{m}_{\bm k}\bigg) n_{FD}(\epsilon^m_{\bm k}) + \bm{\beta}^{m}_{\bm k} n^{(-1)m}_{E{\bm k}} \bigg\}
\end{equation}
\end{widetext}

Let us consider the expectation value of the spin operator in the presence of an electric field  
\begin{equation}
\label{ }
{\rm Tr}\{ \bm{s} f \}= \sum_{m,\bm k}\bm{s}^{mm}_{\bm k} 
n^{m}_{E{\bm k}} + \sum_{n,m,\bm k}\bm{s}^{nm}_{\bm k}S^{mn}_{E\bm k},
\end{equation}
where $\bm{s}^{nm}_{\bm k}$ represent the matrix elements of the spin operator. Writing explicitly the off diagonal terms of the density matrix in the average of the spin operator we can separate the intrinsic and extrinsic contributions as
\begin{align}
\label{ }
\langle {\bm s} \rangle^{\rm int} & = \sum_{n,m,\bm k}
\frac{\hbar {\bm s}_{nm}}{i(\epsilon^{m}_{\bm k}-\epsilon^{n}_{\bm k})}
\, D^{mn}_{E\bm k} \\
\label{ }
\langle {\bm s} \rangle^{\rm ext}
& = \sum_{n,m,\bm k}
\frac{\hbar \bm{s}^{nm}_{\bm k}}{i(\epsilon^{m}_{\bm k}-\epsilon^{n}_{\bm k})} \, D^{'mn}_{E\bm k}.
\end{align}
If we define the quantity 
\begin{equation}
\label{Eq:EffectiveInterbandSpin}
{\bm{\mathcal{T}}}^{nm}_{\bm k}= \frac{\hbar }{i(\epsilon^{n}_{\bm k}-\epsilon^{m}_{\bm k})} {\bm{s}}^{nm}_{\bm k},
\end{equation}
the intrinsic contribution can be rewritten as
\begin{align}
\nonumber
\label{ }
\langle {\bm s} \rangle^{\rm int} &=
\frac{ie}{\hbar}
\sum_{m,\bm k}
[\bm{\mathcal{T}}_{\bm k},\bm E\cdot \bm{\mathcal{R}'}_{\bm k} ]^m n_{FD}(\varepsilon^m_{\bm k}), 
\end{align}
where we only have to take off-diagonal components inside the commutator. This enables us to define an intrinsic spin
expectation expectation value as
\begin{align}
{\bm s}^{\rm int}_{\bm k} 
=
\frac{ie}{\hbar}
[\bm{\mathcal{T}}_{\bm k},\bm E\cdot \bm{\mathcal{R}'}_{\bm k} ]^m.
\end{align}
The extrinsic part can be written as
\begin{align}
\label{ }
\langle \bm s\rangle^{\rm ext}
&=\sum_{n,m,\bm k}  
\bm{\mathcal{T}}^{nm}_{\bm k}[J_0(n^{(-1)}_{E})]^{mn}_{\bm k}
\end{align}
This is mathematically analogous to the expression for the extrinsic interband velocity Eq. \eqref{Eq:ExtrinsicAverageVelocity}, and using a similar manipulation we can re-express it as
\begin{align}
\label{ }
\langle \bm s \rangle^{\rm ext}_{\rm bm}
&=
\sum_{m,\bm k} n^{(-1)m}_{E{\bm k}}
{\bm \gamma}^{mm}_{\bm k}
\end{align}
where $n^{(-1)}_{E{\bm k}}$ is the leading order distribution function and we introduce the new extrinsic spin $\bm{\gamma}^{m}_{\bm k}$ given by
\begin{align}
\label{Eq:ExtrinsicaSpinValue}
{\bm \gamma}^{m}_{\bm k}
&=
\frac{1}{2\pi\hbar} 
\int^{\infty}_{-\infty} d\epsilon 
\langle [U,G^A_0(\epsilon)[U,\bm{\mathcal{T}}]G^R_0(\epsilon)]\rangle^{mm}_{\bm k}.
\end{align}
In exact analogy with ${\bm \beta}^{m}_{\bm k}$, since the Born approximation scattering term vanishes when the distribution function is replaced by the Fermi-Dirac distribution, ${\bm \gamma}^{m}_{\bm k}$ also vanishes in equilibrium in the presence of scalar scattering. This quantity represents spin rotations during scattering events. Again, the quantity $\bm{\mathcal{T}}$ only has inter-band matrix elements. Explicitly $\bm{\gamma}^{m}_{\bm k}$ is evaluated as
\begin{widetext}
\begin{align}
{\bm \gamma}^{mm}_{\bm k}
&=
\frac{\pi}{\hbar}
\sum_{n,m',\bm k'} 
\Big\{
\left[
\bm{\mathcal{T}}^{m,n}_{\bm k}
\langle
U^{nm'}_{\bm k,\bm k'}
U^{m'm}_{\bm k',\bm k}
\rangle
+
\langle
U^{mm'}_{\bm k,\bm k'}
U^{m'n}_{\bm k',\bm k}
\rangle
\bm{\mathcal{T}}^{n,m}_{\bm k}
\right]
\delta(\epsilon^{m}_{\bm k}-\epsilon^{m'}_{\bm k'}) \\ \nonumber
&-
\left[
\langle U^{mn}_{\bm k,\bm k'} U^{m'm}_{\bm k',\bm k} \rangle \bm{\mathcal{T}}^{n,m'}_{\bm k'} 
+
\langle U^{mm'}_{\bm k,\bm k'} U^{nm}_{\bm k',\bm k} \rangle \bm{\mathcal{T}}^{m',n}_{\bm k'} 
\right] 
\delta(\epsilon^{n}_{\bm k'}-\epsilon^{m}_{\bm k})\Big\}.
\end{align}
\end{widetext}

In analogy to the above, we can write a modified intra-band spin matrix element as
\begin{align}
\label{Eq:ModifiedSpinExpectationValue}
{\bm s}^{mm} 
\rightarrow
\bm{s}^{mm}_{\bm k} +
\frac{ie}{\hbar}
[\bm{\mathcal{T}}_{\bm k},\bm E\cdot \bm{\mathcal{R}'}_{\bm k} ]^m
+
\bm{\gamma}^{m}_{\bm k}.
\end{align}
The first factor is the bare matrix element of the spin operator in band $m$, while the remaining two represent non-equilibrium corrections, the first being intrinsic and the second extrinsic. We note that $\bm \gamma_{\bm k}$ and $\bm \beta_{\bm k}$ are mathematically very similar and both contribute at the Fermi energy. The intrinsic contribution to the spin density from the Fermi sea appears in the second term in Eq. \eqref{Eq:ModifiedSpinExpectationValue}. The full semiclassical expression for the spin density is given by
\begin{widetext}
\begin{equation}
\langle \bm s \rangle 
=  \, \sum_{m{\bm k}} 
\bigg\{ 
\bm s^m_{\bm k}
[n_{FD}(\varepsilon^m_{\bm k}) + n^{(-1)m}_{E{\bm k}} + n^{(sk)m}_{E{\bm k}} + n^{(sj)m}_{E{\bm k}}] 
+ 
\frac{ie}{\hbar}
[\bm{\mathcal{T}}_{\bm k},\bm E\cdot \bm{\mathcal{R}'}_{\bm k}]^m
n_{FD}(\epsilon^m_{\bm k}) 
+ 
\bm{\gamma}^{m}_{\bm k} n^{(-1)m}_{E{\bm k}} \bigg\}
\end{equation}
\end{widetext}

\section{Linear response family tree}
\label{Sec:FamilyTreeLRT}

\begin{figure}[t!]
	\includegraphics[width=\linewidth]{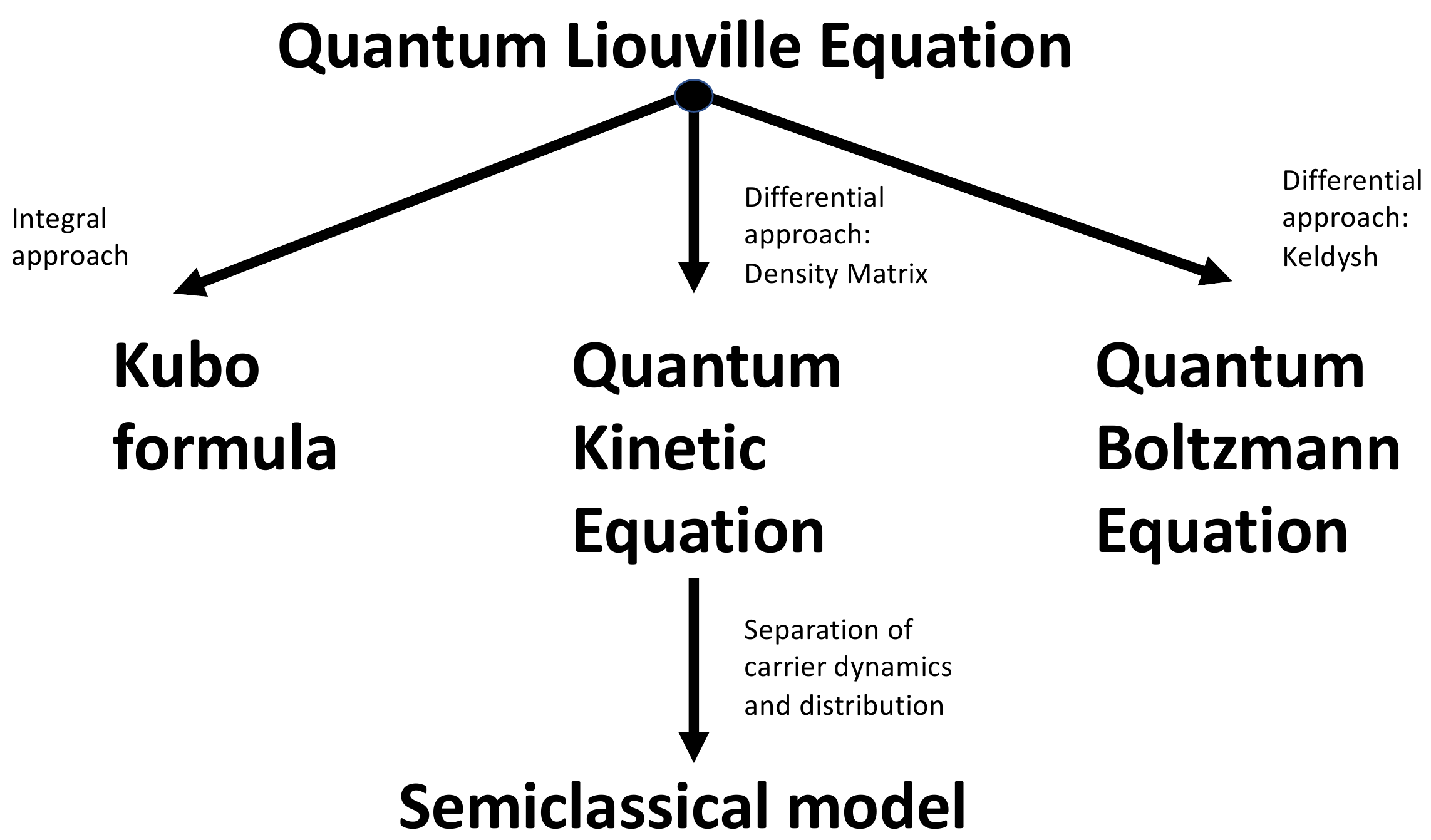}
	\caption{Family tree of linear response theories.}
	\label{FamilyTree}
\end{figure}

In closing the methodological discussion we remark briefly on the relationships between the various linear response theories. The most common strategy for solving Eq.~(\ref{Liouville}) in an electric field is via Kubo linear response theory, which is discussed in detail in many textbooks\cite{HenrikBruss2004}, hence we only dwell upon its fundamental aspects. Briefly, the Hamiltonian is decomposed as $H = (H_0 + U) + V$, and the non-equilibrium part of the density matrix is likewise singled out as $\rho = \rho_0 + \rho_E$. Then in linear response one can write
\begin{equation}\label{K1}
\frac{\partial \rho_E}{\partial t} + \frac{i}{\hbar}\, [H_0 + U,\rho_E] = - \frac{i}{\hbar} \, [V, \rho_0].
\end{equation}
This equation is solved immediately to yield
\begin{equation}\label{rho_Kubo}
\rho_E = \frac{1}{2\pi i}\int^{\infty}_{0}d\epsilon [G^{R}(\epsilon)V G^{A}(\epsilon) , \rho_0],
\end{equation}
where $G^R$, $G^A$ are the retarded and advanced Green's functions, respectively, for the disordered system:
\begin{align}
\label{}
G^R(\epsilon)
&=-\frac{i}{\hbar}\int^{\infty}_{0} dt e^{-i(H_0 + U)t/\hbar} e^{i\epsilon t/\hbar} e^{-\eta t}.
\end{align}
We refrain from writing out the energy dependence in full. Note that the Green's functions have not been averaged over disorder configurations at this stage. To obtain the customary Kubo formula one must trace over the velocity operator and average over impurity configurations
\begin{equation}\label{Kubo}
\bkt{\bm j} = -\frac{e}{2\pi i}\int^{\infty}_{0}d\epsilon \, {\rm tr} \, \bkt{{\bm v} [G^{R}V G^{A} , \rho_0]}.
\end{equation}
The procedure is standard, so we do not cover it here in detail, the purpose of this description is illustrative. The important point to notice is that the Kubo formula is the integral approach to solving the Liouville equation, whereas the kinetic equation we follow in this work represents the differential approach, or integro-differential in view of the complex scattering term. The Keldysh theory follows a similar path, and although it takes as its starting point a series of Green's functions, its ultimate origin lies in the quantum Liouville equation. The Keldysh theory is formally non-local in time, although in the vast majority of practical applications the non-locality is removed and the Keldysh Green's function, which is analogous to the density matrix employed in this work, depends only on the difference in time variables. Thus, for the purposes of the present comparison, the quantum Boltzmann equation derived in the Keldysh theory is indistinguishable from the quantum kinetic equation derived from the density matrix. The Kubo formula, Keldysh theory and quantum kinetic equation may be regarded as \textit{holistic} approaches, in which both the carrier dynamics and the carrier distribution are accounted for in the density matrix (or Keldysh Green's function), and the net result is the expectation value of a physical observable. In contrast, the semiclassical theory involves a separation between the carrier dynamics and carrier distribution, which can help to build an intuitive picture of the underlying physics. The relationship between the different approaches is summarized in the \textit{family tree} of Fig.~\ref{FamilyTree}. The most important observation in this context is the common origin of all linear response theories, which reinforces the expectation that they should all lead to the same results.  

\section{Applications}
\label{Sec:Applications}

We now turn to applications of the theory, which are intended to illustrate the way the extrinsic interband velocity, extrinsic spin terms, and additional scattering terms in the Boltzmann equation appear in the explicit evaluations of physical observables for model systems. In particular, we emphasize the relationship between these various contributions and the analogous quantities appearing in diagrammatic theories, which enables us to reconcile the semiclassical and diagrammatic results. Our focus will be on topological insulators\cite{Culcer2012}, where we discuss the anomalous Hall effect as well as spin-orbit torques, and compare the semiclassical results with previous work. 

\subsection{Anomalous Hall effect in topological insulators}

In this section we calculate the anomalous Hall conductivity in topological insulators. The anomalous Hall conductivity is basically expressed in terms of four contribution: intrinsic contribution 
that takes into account the whole Fermi sea of the system, extrinsic contribution due to the extrinsic velocity $\bm \beta^{m}_{\bm k}$ at the Fermi 
surface, side jump like contribution at the Fermi energy due to an electric field correction to the collision integral and a skew scattering 
contribution.

The Hamiltonian that describes low energy excitations in the surface of 3D topological insulators reads:
\begin{equation}
\label{Eq:effectiveHamiltonian}
H=\hbar v_F (k_x \sigma_y-k_y\sigma_x)+M\sigma_z,
\end{equation}
where $v_F$ is the effective Fermi velocity and $\sigma_{i}$ are Pauli matrices, while $M$ is the magnetization. The eigenvalues are $\epsilon^{\pm}_{\bm k}=\pm\sqrt{\hbar^2v^2_Fk^2+M^2}$, where $\pm$ labels conduction/valence band and the eigenstates are
\begin{equation}
\label{Eq:EigenStates}
|u^{\pm}_{\bm k}\rangle=
\frac{1}{\sqrt{2}}
\begin{pmatrix}
e^{-i\theta_{\bm k}} \sqrt{1\pm\xi_k}\\
\pm i  \sqrt{1\mp \xi_k } 
\end{pmatrix},
\end{equation}
with a parameter $\xi_k=M/\lambda_{\bm k}$ with $\lambda_{\bm k}=\sqrt{\hbar^2v^2_Fk^2+M^2}$ and $\theta_{\bm k}=\arctan(k_y/k_x)$. 

From the eigenstates we determine the Berry connection vector $\bm{\mathcal{R}}^{mm'}_{\bm k}=i\langle u^{m}_{\bm k}|\nabla_{\bm k}u^{m'}_{\bm k}\rangle$. We decompose this vector into a diagonal and an off-diagonal contribution, namely, 
$\bm{\mathcal{R}}_{\bm k}=\bm{\mathcal{R}}^{d}_{\bm k}+\bm{\mathcal{R}}^{od}_{\bm k}$, with
\begin{align}
\label{}
\bm{\mathcal{R}}^{d}_{\bm k}
&=\frac{\hat{\bm \theta } }{2k}(\sigma_0+\xi_{\bm k}\sigma_z) \\
\bm{\mathcal{R}}^{od}_{\bm k}
&=\frac{\hbar v_F}{2\lambda_{\bm k}}(\sigma_x\hat{\bm \theta } - \xi_{\bm k} \sigma_y\hat{\bm k}). 
\end{align}

We have defined a unit vector parallel to momentum 
$\hat{\bm k}=(\cos\theta_{\bm k},\sin\theta_{\bm k})$ 
and a unit vector perpendicular to momentum 
$\hat{\bm \theta} =(-\sin\theta_{\bm k},\cos\theta_{\bm k})$. Also, these vectors are related as $\hat{\bm \theta }=\hat{\bm z}\times\hat{\bm k}$.


Let us start by calculating the intrinsic contribution to the Hall conductivity. It reads:
\begin{equation}
\label{ }
\sigma^{\rm int}_{yx} = \frac{e^2}{\hbar}
\sum_{m,\bm k} \Omega^{m}_{\bm k,z} n_{FD}(\epsilon^{m}_{\bm k}). 
\end{equation}
Explicit evaluation for a topological insulator gives the Berry curvature 
\begin{equation}
\label{Eq:BerryCurvature}
\Omega^{\pm}_{k,z}
=
\mp   
\frac{\xi_{k}(1-\xi^2_{k})}{2k^2}.
\end{equation}
After a straightforward integration we find the intrinsic contribution to the Hall conductivity to be
\begin{align}
\label{ }
\sigma^{\rm int}_{yx}
&=
\frac{e^2}{4\pi  \hbar} 
\frac{M}{\epsilon_{F}}. 
\end{align}
The extrinsic inter-band velocity makes the following contribution to the anomalous Hall conductivity
\begin{align}
\label{Eq:Extrinsic}
\sigma^{\rm ext}_{yx}
&=-\frac{e}{E_x}\sum_{m,\bm k} n_{E{\bm k}}^{(-1)m} \beta^{m}_{\bm k,y}.
\end{align}
The leading order correction to the distribution function is $n_{E{\bm k}}^{(-1)m} =-e\tau_{tr} \bm E \cdot \bm v^{m}_{\bm k}\delta(\epsilon^{m}_{\bm k}-\epsilon_F)$ with the transport time given by the expression 
\begin{align}
\label{ }
\frac{1}{\tau_{tr}}=\frac{1}{2\tau}\left( 1+3\xi^2_{\bm k}\right),
\end{align}
with the scattering time defined as $1/\tau=\pi u^2_0\rho(\epsilon_{\bm k})/\hbar$ and the density of states $\rho(\epsilon_{\bm k})=\lambda_{k}/2\pi \hbar^2v^2_F$. The diagonal velocity is $v_x=v_F\left( 1-\xi^2_{\bm k} \right)^{1/2}\cos\theta_{\bm k}$ and the extrinsic inter-band velocity in the conduction band takes the form 
\begin{align}
\label{ }
\bm \beta
&=
\sigma_0\frac{1}{\tau}\frac{\hbar v_F}{\lambda_{k}}\xi_{k}(1-\xi^2_{k})^{1/2}\hat{\bm \theta }.
\end{align}

The extrinsic inter-band velocity is a transverse velocity since it is proportional to the unit vector $\hat{\bm \theta}$. After explicit integration we arrive at the expression 
\begin{align}
\label{Eq:ExtrinsicHallEffect}
\sigma^{\rm ext}_{yx}
&=
\frac{e^2}{2\pi \hbar} 
\frac{M}{\epsilon_F}
\frac{\left( 1-\xi^2_{F} \right)}{\left(1+3\xi^2_{F} \right)}. 
\end{align}

Notice that $n_{E{\bm k}}^{(-1)m}$ is inversely proportional to the impurity density while the extrinsic velocity $\bm \beta$ is proportional to the impurity density. As a result, the overall effect in Eq. \eqref{Eq:Extrinsic} is independent of disorder. The extrinsic inter-band velocity $\bm{\beta}^{m}_{\bm k}$ comprises the effect of disorder on carrier dynamics, namely, it can be interpreted as an effective velocity of the electron after many collisions, in contrast to the group velocity, which is a velocity between collisions. In this sense our extrinsic velocity can be associated to the side jump velocity encountered in previous semiclassical results \cite{Sinitsyn2006_II} but with the difference that $\bm{\beta}^{m}_{\bm k}$ is entirely due to band mixing mediated by the off-diagonal components of the Berry connection vector and that it is constructed from a collision integral without introducing any quantity related to coordinate shift.  

As we discussed earlier, there are also two contributions to the anomalous Hall conductivity related to two different diagonal subleading density matrix functions. Let us first calculate the anomalous Hall conductivity related to the term $n^{(\rm sj)}_{E {\bm k}}$ in the distribution function. It reads
\begin{align}
\sigma^{\rm sj}_{yx}
&=
-\frac{e}{E_x}
\sum_{\bm k}
v^{++}_{\bm k,y} n^{(\rm sj)++}_{E {\bm k}}.
\end{align}
The diagonal velocity reads
\begin{align}
\label{ }
v^{++}_{\bm k,y}&=v_F(1-\xi^2_{\bm k})^{1/2} \sin\theta_{\bm k},
\end{align}
while the correction to the distribution function is
\begin{align}
\label{Eq:subleadingdensityfunctionsidejump}
n_{E{\bm k}}^{(\rm sj)++} 
&=
2\frac{\hbar v_F}{\lambda_{k}}
\xi_{k} 
\delta(\epsilon_F-\epsilon^{+}_{\bm k})
\frac{(1-\xi^2_{k})^{1/2}}{(1+3\xi^2_{\bm k})}
e\bm E \cdot \hat{\bm \theta }. 
\end{align}
Replacing all elements we find for the conductivity 
\begin{align}
\label{Eq:SideJumpConductivity}
\sigma^{\rm sj}_{yx}
&=
\frac{e^2}{2\pi\hbar}
\frac{M}{\epsilon_F}
\frac{(1-\xi^2_{F})}{(1+3\xi^2_{F})}.
\end{align}
This term doubles the contribution in Eq.\eqref{Eq:ExtrinsicHallEffect} due to the extrinsic velocity although in this case the effect of disorder is completely captured by $n^{(\rm sj)}_{E{\bm k}}$. 

In previous semiclassical studies an anomalous distribution function was introduced as a result of a coordinate shift \cite{Sinitsyn2006_II}. In contrast, we have derived $n^{(\rm sj)}_{E{\bm k}}$ from Eq. \eqref{Eq:balanceEquationSideJump} without any need for introducing a coordinate shift. 

The contribution to the anomalous Hall conductivity related to the skew scattering correction to the distribution function, $n^{(\rm sk)}_{E {\bm k}}$, reads
\begin{align}
\sigma^{\rm sk}_{yx}
&=
-\frac{e}{E_x}\sum_{\bm k}
\left[v^{++}_{\bm k,y} n^{(\rm sk)++}_{E {\bm k}}+v^{--}_{\bm k,y} n^{(\rm sk)--}_{E {\bm k}}\right].
\end{align}
The diagonal velocities are 
$ v^{\pm \pm}_{\bm k,y}= \pm v_F(1-\xi^2_{\bm k})^{1/2} \sin\theta_{\bm k}$, and
\begin{align}
\label{Eq:SubleadingDensityFunctionSkew}
n_{E{\bm k}}^{(\rm sk)}
&=
\frac{3}{2}
\frac{\hbar v_F}{\lambda_{k}}
\xi_{k}
\frac{\left( 1-\xi^2_{F} \right)^{3/2}}{\left(1+3\xi^2_{F} \right)^2}
\delta(\epsilon_{F}-\epsilon^{+}_{\bm k})
e\bm E\cdot \hat{\bm \theta }\sigma_z.
\end{align}
After integration we obtain
\begin{align}
\label{Eq:SkewScatteringConductivity}
\sigma^{\rm sk}_{yx}
&=
\frac{e^2}{2\pi \hbar}
\frac{M}{\epsilon_{F}}
\frac{3(1-\xi^2_{F})^{2}}{2(1+3\xi^2_{F})^{2}}. 
\end{align}
Adding all the contributions to the Hall conductivity we get the final expression
\begin{align}
\label{Eq:TotalHallConductivity}
\sigma_{yx}
&=\frac{4e^2}{2 \pi \hbar} 
\frac{M}{\epsilon_{F}}
\left[
\frac{1+\xi^2_{F}}{(1+3\xi^2_{F})^{2}}
\right],
\end{align}
in exact agreement with previous results using the non-crossing approximation and diagrammatic perturbation theory \cite{Sinitsyn2006_II, Sinitsyn_2007, Sinitsyn_2007JPCM, Ado_2015,Nagaosa2010}.

\subsection{Spin density and spin-orbit torques in topological insulators}

In this section we determine the spin density and spin-orbit torques in topological insulators with an out-of-plane magnetization described by the effective Hamiltonian Eq. \eqref{Eq:effectiveHamiltonian}. As for the conductivity, the spin density has five contributions: a dominant contribution from the Fermi surface leading to the Edelstein effect\cite{Edelstein1990}, an intrinsic contribution from the Fermi sea, a contribution due to the extrinsic spin expectation value $\bm \gamma^m_{\bm k}$ at the Fermi surface, a side-jump like contribution at the Fermi energy due to the electric field correction to the collision integral and a skew scattering 
contribution.

The leading order contribution to the spin density is 
\begin{equation}
\langle {\bm s} \rangle^{\rm Edel}
= \sum_{m,\bm k}{\bm s}^{mm}_{\bm k} n_{E{\bm k}}^{(-1)m}.
\end{equation}
Using $n_{E{\bm k}}^{(-1)m} =-e \tau_{tr} {\bm E} \cdot \bm v^{m}_{\bm k}\delta(\epsilon^{m}_{\bm k}-\epsilon_F)$ and
\begin{align}
\label{ }
s^{\pm \pm}_{\bm k, x}&=\mp(1-\xi^2_{\bm k})^{1/2} \sin\theta_{\bm k} \\ 
s^{\pm \pm }_{\bm k, y}&=\pm (1-\xi^2_{\bm k})^{1/2}\cos\theta_{\bm k},
\end{align}
the Edelstein effect contribution to the spin density is 
\begin{equation}
\langle {\bm s} \rangle^{\rm Edel} = e\tau {\bm E} \times \hat{\bm z} v_F \rho(\epsilon_{F}) 
\frac{1-\xi^2_{F}}{1+3\xi^2_{F}}.
\end{equation}

The fraction of the spin density weighted by the intrinsic driving term reads:
\begin{align}
\label{ }
\langle {\bm s} \rangle^{\rm int}
&= \frac{ie\bm E}{\hbar}\cdot
\sum_{nm\bm k}
(
\bm{\mathcal{T}}^{mn}_{\bm k}
\bm{\mathcal{R}}^{nm}_{\bm k}
-
\bm{\mathcal{R}}^{mn}_{\bm k}
\bm{\mathcal{T}}^{nm}_{\bm k}
) n_{FD}(\varepsilon^m_{\bm k}).
\end{align}
The dot product is between the electric field and the Berry connection vector. With the Berry connection 
\begin{align}
\label{}
\mathcal{R}^{+-}_{\bm k,x}
&=
-\frac{\hbar v_F}{2\lambda_{\bm k}}(\sin\theta_{\bm k}- i\xi_{\bm k}\cos\theta_{\bm k}) \\
\mathcal{R}^{+-}_{\bm k,y} &=
\frac{\hbar v_F}{2\lambda}(\cos\theta_{\bm k}+i\xi_{\bm k}\sin\theta_{\bm k})
\end{align}
and the off-diagonal spin expectation values
\begin{align}
\label{ }
s^{-+}_{\bm k, x} &= (\xi_{\bm k} \sin\theta+i\cos\theta) \\ 
s^{-+}_{\bm k, y} &= -(\xi_{\bm k}\cos\theta -i \sin\theta  )  
\end{align}
we get the intrinsic correction to the spin density 
\begin{align}
\label{ }
\langle {\bm s} \rangle^{\rm int} &= \frac{eM\hbar v_F\rho(\epsilon_F)}{2\epsilon^2_F} \, {\bm E}.
\end{align}

The extrinsic correction is defined as
\begin{align}
\langle {\bm s} \rangle^{\rm ext} & = \sum_{m,\bm k} n_{E{\bm k}}^{(-1)m} {\bm \gamma}^{m}_{\bm k} \\ 
\bm{\gamma}^{++}_{\bm k}
&=
-\sigma_0\frac{1}{\tau}
\frac{\hbar}{\lambda_{\bm k}} 
\xi_{k}(1-\xi^2_{k})^{1/2}\hat{\bm k},
\end{align}
yielding
\begin{align}
\label{Eq:SpinDensityExtrinsicEx}
\langle {\bm s} \rangle^{\rm ext}
= \frac{e \hbar v_F\rho(\epsilon_{F}) M}{\epsilon^2_{F}}
\frac{1-\xi^2_{F}}{1+3\xi^2_{F}} \, {\bm E}.
\end{align}
This extrinsic contribution to the spin density is the counterpart of the extrinsic inter-band velocity contribution in the anomalous Hall effect. The extrinsic spin $\bm{\gamma}^{m}_{\bm k}$ as defined in Eq. \eqref{Eq:ExtrinsicaSpinValue} is an interband coherence effect mediated by an effective off-diagonal spin operator defined in Eq. \eqref{Eq:EffectiveInterbandSpin}.

The side-jump contribution is given by
\begin{align}
\langle s_x \rangle^{\rm sj}
&=
\sum_{\bm k}s^{++}_{\bm k,x} n^{(\rm sj)++}_{E{\bm k}}.
\end{align}
Using the spin expectations values above and the density function given by Eq.\eqref{Eq:subleadingdensityfunctionsidejump} we find that
the side-jump contribution to the spin density reads 
\begin{align}
\langle {\bm s} \rangle^{\rm sj}
& = \frac{e\hbar v_F
\rho(\epsilon_{F}) M}{\epsilon^2_{F}}
\frac{(1-\xi^2_{F}) }{(1+3\xi^2_{F})} \, {\bm E}.  
\end{align}
This term doubles the contribution of
Eq. \eqref{Eq:SpinDensityExtrinsicEx}. This is the counterpart of the anomalous distribution function introduced in the semiclassical theory \cite{Sinitsyn2006_II} and also calculated in Eq. \eqref{Eq:SideJumpConductivity} for the anomalous Hall conductivity.

The skew scattering contribution takes the form
\begin{align}
\langle {\bm s} \rangle^{\rm skew}
&=
\sum_{\bm k}\left[{\bm s}^{++}_{\bm k,x} n^{(\rm sk)++}_{E{\bm k}} + {\bm s}^{--}_{\bm k} n^{(\rm sk)--}_{E{\bm k}}\right].
\end{align}
Using the diagonal spin expectations values and the distribution function found in Eq.\eqref{Eq:SubleadingDensityFunctionSkew} we find for the skew scattering contribution to the spin density 
\begin{align}
\langle {\bm s} \rangle^{\rm sk}
&= \frac{e\hbar\rho(\epsilon_F) v_F\xi_{F}}{\lambda_{F}} \frac{3(1-\xi^2_{F})^{2}}{2(1+3\xi^2_{F})^{2}} \, {\bm E}.
\end{align}
This is the counterpart of the skew scattering term in the anomalous Hall conductivity calculated in Eq. \eqref{Eq:SkewScatteringConductivity}.

Finally, the total spin density of the system then reads
\begin{align}
\label{Eq:TotalSpinDensity}
\bkt{\bm s}
&=
-e\tau v_F
\rho(\epsilon_{F})
\frac{(1-\xi^2_{F})}{\left( 1+3\xi^2_{F}\right)} \hat{\bm z}\times \bm E 
\\ \nonumber
&+
4e \hbar v_F
\rho(\epsilon_{F})
\frac{M}{\epsilon^2_{F}}
\frac{(1+\xi^2_{F})} {(1+3\xi^2_{F})^{2}} \, {\bm E}. 
\end{align}
The spin-orbit torque is defined as $\bm \tau=(2M/\hbar)\bm m \times \langle \bm s \rangle$, where $\bm m$ is a unit vector in the direction of magnetisation that we took here as $\bm m=\hat{\bm z}$. Restoring the factor of $\hbar/2$ in the spin matrix elements and substituting the density of states $\rho(\epsilon_F)=\epsilon_F/2\pi \hbar^2v^2_F$, the spin-orbit torque is finally given by
\begin{align}
\label{Eq:spinorbittorque}
\bm {\tau}
&= - \frac{e\tau \epsilon_FM
\rho(\epsilon_{F})}{2\pi \hbar^2 v_F} \frac{(1-\xi^2_{F})}{\left( 1+3\xi^2_{F}\right)} \bm m \times (\hat{\bm z}\times \bm E) \\ \nonumber 
& +
\frac{2e}{\pi \hbar v_F}
\frac{M^2}{\epsilon_{F}}
\frac{(1+\xi^2_{F})} {(1+3\xi^2_{F})^{2}} {\bm m} \times {\bm E}, 
\end{align}
in agreement with previous results. \cite{Ndiaye_2017,Sakai2014} A peculiarity of topological insulators is that the velocity operator is directly related to the spin operator as 
$\hat{\bm v}=-v_F\hat{\bm z}\times \bm \sigma$. Then the current density is $\bkt{\bm j} =-e\langle \hat{\bm v} \rangle$
what implies that the anomalous Hall conductivity and the spin density are related by $\sigma_{yx}=J_{y}/E_x=ev_F\langle s_x \rangle$. This fact can indeed be verified from Eq. ~\eqref{Eq:TotalHallConductivity} and Eq. ~\eqref{Eq:TotalSpinDensity}. 

\section{Conclusions and Outlook}
\label{Sec:Conclusion}

We have demonstrated that the semiclassical dynamics of electrons in disordered solids can be determined using linear response theory by taking the density matrix and quantum Liouville equation as the starting point. This results in a disorder-dependent correction to the semiclassical equation of motion for the carrier position, which we have termed the \textit{extrinsic inter-band velocity} $\bm{\beta}^{m}_{\bm k}$, and which accounts for the effect of disorder on the carrier velocity after many collisions. In analogy to the extrinsic inter-band velocity, an extrinsic correction to the spin expectation value $\bm{\gamma}^{m}_{\bm k}$ is also present, which accounts for spin rotations during scattering events. This is accompanied by an intrinsic, electric field dependent non-equilibrium correction to the spin expectation value, which is analogous to the \textit{anomalous velocity} present in the semiclassical equations of motion. At the same time, the Boltzmann equation must be solved up to sub-leading order in the disorder strength. The collision integral in the Boltzmann equation includes an electric field dependent correction which is analogous to the \textit{side-jump} scattering term, as well as an additional correction analogous to the \textit{skew scattering} term in systems with intrinsic spin-orbit interactions. We have applied this theory to describe the anomalous Hall effect and spin-orbit torques in topological insulators, obtaining exact agreement with quantum mechanical results using the Kubo formula. 
 
The general prescription we have formulated in this work can be straightforwardly generalised to include extrinsic spin-orbit and magnetic impurity scattering, as well as magnetic fields, which, however, require more effort since the description relies on the Wigner function. Our prescription paves the way towards a systematic semiclassical picture encompassing a host of effects that conventionally lie beyond the purview of semiclassical theory: disorder effects beyond the Born approximation, such as weak localization, electron-electron interactions in the mean-field approximation, as well as Kondo physics in magnetic systems.

\textit{Acknowledgments}. DC is supported by the Australian Research Council Future Fellowship FT190100062. We would like to thank Leonid Glazman, Aydin Keser and Dmitry Efimkin for a series of educational discussions. 



\bibliography{SD}

\end{document}